\documentclass[hyper]{JHEP3}
\usepackage{epsfig}
\usepackage{subfigure}
\usepackage{amsfonts}
\usepackage{latexsym}
\usepackage{amsmath}
\usepackage{cite}

\def\half{{\textstyle{1\over2}}}

\def\e{{\epsilon}}

\def\p{\partial}

\def\r{\rightarrow}
\def\om{\omega}

\def\s{\sigma}

\def\a{\alpha}
\def\b{\beta}
\def\l{\lambda}

\def\tr{\tilde{r}}
\def\th{\theta}
\def\tL{\tilde{L}}
\def\O{\mathcal{O}}
\def\d{\delta}

\newcommand{\bea}{\begin{eqnarray}}
\newcommand{\eea}{\end{eqnarray}}
\newcommand{\be}{\begin{equation}}
\newcommand{\ee}{\end{equation}}
\newcommand{\non}{\nonumber}
\newcommand{\bi}{\begin{itemize}}
\newcommand{\ei}{\end{itemize}}
\title{{Stability of warped $AdS_3$ vacua of \\ topologically massive gravity} }
\preprint{}
\author{
Dionysios Anninos$^{1}$, Mboyo Esole$^{1}$ and Monica Guica$^{2}$\\

\it $^1$ Jefferson Physical Laboratory, Harvard University, \\
\it 17 Oxford St., Cambridge, MA 02138, USA\\

\it $^2${Laboratoire de Physique Th\'eorique et Hautes Energies (LPTHE)\\
\it{Universit\'e Pierre et Marie Curie-Paris 6; CNRS UMR 7589}\\
\it{Tour 24-25, 5$^{\grave{e}me}$ \'etage, Boite 126, 4 Place Jussieu} \\
\it {75252 Paris Cedex 05, France}}\\

}

\abstract{
\vskip 3 mm
$AdS_3$ vacua of topologically massive gravity (TMG) have been shown to be perturbatively unstable for all values of the coupling constant except the chiral point $\mu \ell=1$. We study the possibility that the warped vacua of TMG, which exist for all values of $\mu$, are stable under linearized perturbations. In this paper, we show that spacelike warped $AdS_3$ vacua with Comp\`{e}re-Detournay boundary conditions are indeed stable in the range $\mu\ell > 3$. This is precisely the range in which black hole solutions arise as discrete identifications of the warped $AdS_3$ vacuum. The situation somewhat resembles chiral gravity: although negative energy modes do exist, they are all excluded by the boundary conditions, and the perturbative spectrum solely consists of boundary (pure large gauge) gravitons. }


\begin{document}

\section{Introduction}

Topologically massive gravity (TMG) \cite{Deser:1981wh,Deser:1982vy} is an interesting extension of three-dimensional gravity which contains  both propagating degrees of freedom as well as black hole solutions. The action of TMG is obtained by adding to the usual Einstein-Hilbert action with a positive Newton constant a gravitational Chern-Simons contribution, with coupling constant $1/\mu$. As for the usual Einstein-Hilbert action, TMG may also be supplemented with a negative cosmological constant $-1/\ell^2$.

A particular vacuum of TMG with a negative cosmological constant is $AdS_3$, which also contains the BTZ black holes \cite{Banados:1992gq,Banados:1992wn}. For arbitrary Chern-Simons coefficient, the $AdS_3$ vacuum suffers from perturbative instabilities. However, it was noted in \cite{Li:2008dq} that at the special point $\mu\ell=1$, the $AdS_3$ vacuum is stable and the theory has a purely chiral spectrum \cite{Strominger:2008dp}. It has been shown in \cite{Maloney:2009ck} that the quantum partition function of chiral gravity has all the required features for the theory to be dual to an extremal CFT \cite{Witten:2007kt}.

$AdS_3$ is only one of various possible vacua of TMG with a cosmological constant. As shown in \cite{Vuorio:1985ta,Percacci:1986ja,Ortiz:1990nn,Nutku:1993eb,Gurses}, less symmetric vacua known as warped $AdS_3$ occur as classical solutions to the equations of motion. These solutions are specific to TMG, i.e. they are not solutions of pure Einstein gravity with a cosmological constant. Their defining property is that they are real line fibrations over $AdS_2$ preserving a single $SL(2,\mathbb{R})$ isometry of the original $SL(2,\mathbb{R})_L \times SL(2,\mathbb{R})_R$ $AdS_3$ isometries, along with a non-compact $U(1)$ isometry generated by translations along the fibre coordinate.

The warped vacua of TMG fall into three types: spacelike, timelike and null warped, depending on whether the norm of the Killing vector generating the $U(1)$ isometry is positive, negative or zero. Each of the first two types can be further classified as  stretched ($\mu\ell>3$) or squashed ($\mu\ell<3$) depending on the magnitude of the warp factor. The fact that these background spacetimes are not asymptotically $AdS_3$ makes them very interesting to study, since we could hope to develop new types of holographic correspondences. In \cite{Anninos:2008fx} such a correspondence was proposed and the central charges of the putative CFT were conjectured\footnote{Related questions have been tackled in the case of the Kerr/CFT correspondence \cite{Guica:2008mu}, and in the  recent subject of theories dual to non-relativistic CFTs \cite{Son:2008ye,Balasubramanian:2008dm}. Warped $AdS_3$ has also been studied in the context of string theory \cite{Anninos:2008qb,Compere:2008cw,Detournay:2005fz,Israel:2004vv,Israel:2003ry}.}.

Quotients of warped $AdS_3$ along various Killing directions may give rise to black holes \cite{Anninos:2008fx}, in perfect analogy with the BTZ case in $AdS_3$. Black hole solutions free of closed timelike curves (CTCs) can only be found in spacelike stretched and null warped $AdS_3$. One can also consider quotients of spacelike warped $AdS_3$ along the $U(1)$. Such geometries have Killing horizons and no CTCs, they resemble the self-dual solutions in $AdS_3$ \cite{Coussaert:1994tu}. Thus, the spacelike warped TMG vacua seem the most interesting to study.

The subject of the present article is the classical stability of these spacetimes. Special attention will be given to the spacelike stretched case, which is richer and better understood. The stability about a certain background depends on the selection of consistent boundary conditions. For example, the propagating mode of TMG in an $AdS_3$ background carries negative energy, thus rendering the theory unstable. Chiral gravity is spared because the massive graviton disappears from the spectrum at $\mu \ell=1$, for Brown-Henneaux boundary conditions \cite{Brown:1986nw,Hotta:2008yq}. There exist several consistent choices of boundary conditions for TMG in $AdS_3$ \cite{Grumiller:2008qz,Henneaux:2009pw}, but only the more restrictive ones exclude the negative energy modes.

We find a similar situation for propagating modes in spacelike stretched  $AdS_3$. While the massive gravitons of warped $AdS_3$ have negative energy, we will see that they do not obey the Comp\`ere-Detournay boundary conditions \cite{Compere:2008cv,Compere:2007in}, which are the only consistent set of boundary conditions proposed in the literature. Thus we discard the massive gravitons from the physical spectrum. These boundary conditions are still relaxed enough to allow the stretched $AdS_3$ black holes, which was the original reason they were
studied.

Having discarded the propagating modes from the spectrum, all we are left with are pure (large) gauge modes. It is well known that if these excitations fall off slowly enough near the boundary of the spacetime they should be included in the physical phase space.  The analysis of the asymptotic symmetry group in spacelike stretched $AdS_3$ performed in \cite{Compere:2008cv} shows that the energy of such pure large gauge modes has to be positive. Incidentally, for our proposed boundary conditions the only remaining excitations have definite chirality. This is quite reminiscent of what happens in chiral gravity.

While we find strong evidence for the stability of stretched $AdS_3$, the case of squashed $AdS_3$ remains inconclusive. The main culprit is our lack of understanding of the boundary of this space and of a set of consistent boundary conditions. Nevertheless, all the explicit and implicit propagating solutions to the linearized TMG equations of motion that we found hold equally well for squashed as they do for stretched $AdS_3$, and they can be used to study the stability of this spacetime once the aforementioned issue is overcome.

The organization of this paper is as follows: in section \ref{back} we review the TMG action, warped $AdS_3$ backgrounds and black holes.  In section \ref{pert} we  describe our general procedure for gauge-fixing, finding the linearized solutions to the TMG equations of motion, imposing boundary conditions and computing the energy density of the gravitational waves. In section \ref{massgrav} we display the explicit highest weight propagating solutions and compute their energy. We also study general propagating solutions. In section \ref{boundgrav} we display  the boundary conditions for stretched $AdS_3$. We conclude with a discussion in section \ref{discussion}. Various derivations and expressions  are presented in the appendices.

\section{Preliminaries I: the background solution}\label{back}

In this section we discuss the basic framework and the background warped $AdS_3$ geometry that we will work with. We also review the black hole solutions obtained from discrete global identifications of the background and the asymptotic structure of these solutions.

\subsection{The action}

The action for topologically massive gravity (TMG) \cite{Deser:1981wh,Deser:1982vy} is
\be
I_{TMG} = \frac{1}{16\pi G}\int d^3x \sqrt{-g}\left[  R + \frac{2}{\ell^2} - \frac{1}{2\mu} \,\varepsilon^{\lambda \mu \nu} \,\Gamma^\rho_{\lambda \sigma} ( \partial_\mu \Gamma^\sigma_{\rho\nu} + \frac{2}{3} \Gamma^\sigma_{\mu\tau} \Gamma^{\tau}_{\nu \rho} ) \right]
\ee
where $\varepsilon^{012} = +1/\sqrt{-g}$ and $\mu$ is a dimensionful coupling with dimensions of mass. While it is well-known that pure three-dimensional gravity possesses no propagating degrees of freedom, the addition of the higher derivative term introduces  a new, massive,  propagating degree of freedom, the so-called massive graviton. When linearizing TMG about flat space, the mass squared of the propagating graviton is $\mu^2$.

It will prove to be convenient to introduce the new quantities $\ell$ and $\nu$, defined as
\be
\Lambda =  -\frac{1}{\ell^2}\;, \;\;\;\;\; \mu = \frac{3 \nu}{\ell}
\ee
In terms of these, the TMG equations of motion are given by
\be
R_{\mu\nu} - \frac{1}{2}R \,g_{\mu\nu} - \frac{1}{\ell^2}g_{\mu\nu} + \frac{\ell}{3\nu}C_{\mu\nu} = 0
\ee
where
\be
C_{\mu \nu} = \half {\varepsilon_{\mu}}^{\alpha \beta}\nabla_{\alpha}(R_{\beta\nu} - \frac{1}{4}g_{\beta\nu}R)
\ee
The Cotton tensor, $C_{\mu\nu}$, is symmetric, traceless and conserved and vanishes on shell for all Einstein solutions.

\subsection{Warped $AdS_3$ backgrounds}

All solutions to pure 3d Einstein gravity with a cosmological constant have vanishing Cotton tensor, so they are automatically solutions of TMG. Nevertheless, there also exist nontrivial solutions particular to TMG, such as the warped $AdS_3$ solutions, which come in several types \cite{Nutku:1993eb,Gurses,Anninos:2008fx,Bouchareb:2007yx}.

We will focus on the vacuum solution known as spacelike warped anti-de Sitter space, with metric:
\be
ds^2 = \frac{\ell^2}{\nu^2+3}\left[ -(1+r^2)\, d\tau^2 + \frac{dr^2}{1+r^2} + \frac{4\nu^2}{\nu^2+3}\,(dx + r d\tau)^2 \right] \label{sqads3}
\ee
where $r,\tau, x\in(-\infty,\infty)$. The boundary of this space resides at $r \r \pm \infty$ for fixed $x$ and $x \r \pm \infty$ for fixed $r$. To simplify our formulae, we will often use the warp factor
\be
a \equiv  \frac{2 \nu}{\sqrt{\nu^2 +3}} \;, \;\;\;\;\;\;\; a\in [0,2)
\ee
If $a>1$ the spacetime is called stretched, whereas if $a<1$  it is called  squashed. Finally, $a=\nu=1$ corresponds to global $AdS_3$.

The above coordinates are geodesically complete, as reviewed in Appendix A. Furthermore, the constant $\tau $ slices are spacelike for all $r$, thus rendering $\tau$ as our global time coordinate. When $\nu^2 \leq 1$, $\partial_\tau$ is a globally defined timelike Killing vector; however, for $\nu^2 > 1$ it becomes spacelike at large $r$. Thus  for $\nu^2 > 1$ our spacetime resembles the region within the ergosphere of a rotating black hole, in the sense that there can be no static observers. It is global identifications of this spacetime with $\nu^2 > 1$ that give rise to black hole solutions with no CTCs \cite{Anninos:2008fx}.

A way to understand the isometries of spacelike warped $AdS_3$ is to notice that for $\nu=1$, the metric \eqref{sqads3} describes $AdS_3$ when written as a Hopf fibration over $AdS_2$
\be
ds^2 = \underbrace{\frac{\ell^2}{4} \left(-(1+r^2)\, d\tau^2 + \frac{dr^2}{1+r^2}\right)}_{AdS_2}  + \underbrace{\frac{\ell^2}{4}(dx + r d\tau)^2}_{fibre}
\ee
The six isometries of $AdS_3$ form the group $SL(2,\mathbb{R})_{L}\times SL(2,\mathbb{R})_{R}$. The $SL(2,\mathbb{R})_{R}$, which is generated by the $\tilde{L}_{\pm1},\tilde{L}_0$ Killing vectors below, leaves the expressions in each of the parentheses invariant. It is then apparent that upon turning on $a \neq 1$, the isometry group of the space will be $SL(2,\mathbb{R})_{R}\times U(1)_{L}$, where $U(1)_L$ - which is noncompact - is generated by $x$-translations. The Killing vectors are
\be
\tilde{L}_0 =   i\, \p_{\tau} \;, \;\;\;\;\; J_0 = - i\, \p_x
\ee
\be
\tilde{L}_{\pm1} = \pm e^{\pm i\tau} \left(\frac{r}{\sqrt{1+r^2}} \,\p_{\tau} \mp i \sqrt{1+r^2} \,\p_r + \frac{1}{\sqrt{1+r^2}} \,\p_x \right)
\ee
and they obey the usual $SL(2,\mathbb{R})$ algebra under Lie brackets:
\be
[\tL_1,\tL_{-1}] = 2 \tL_0 \;, \;\;\;\;\;[\tL_{\pm 1},\tL_0]= \pm  \tL_{\pm 1}\;, \;\;\;\;\; [J_0,\tilde{L}_j] =0, \quad j \in \{0,\pm 1 \}.
\ee

In terms of the original $AdS_3$ Killing vectors, the $SL(2,\mathbb{R})_R$ generators are preserved by the warping, together with the $J_0$ Killing vector of the $SL(2,\mathbb{R})_L$ isometries. The rest are explicitly broken. Note that in spacelike warped $AdS_3$ there is no Killing vector with compact orbits: the isometry generated by $L_0-\tL_0$, which was such a Killing vector in $AdS_3$, is explicitly broken by the warping.

\subsection{Warped $AdS_3$ black holes}\label{wbtz}

As mentioned in the introduction, TMG contains black holes which are locally spacelike stretched $AdS_3$. The metric  is given by
      \begin{multline}
      \frac{ds^2}{\ell^2}=dt^2+\frac{ d\tr^2}{(\nu^2+3)(\tr-r_{+})(\tr-r_{-})} + \left(2\nu \tr  -
      \sqrt{r_{+}r_{-}(\nu^2+3)}\right)dtd\theta \\
      +\frac{\tr}{4}\left(3(\nu^2 - 1)\tr+(\nu^2+3)(r_+ + r_-) - 4\nu\sqrt{r_+r_-(\nu^2+3)}\right)d\theta^2
      \label{ng}
      \end{multline}
where $\theta$ is identified by $2\pi$. The inner and outer horizons are given by $r_-$ and $r_+$ which are positive. Note that the above black holes are free of CTCs only for $\nu \ge 1$. As discussed in \cite{Anninos:2008fx}, these black holes are obtained from discrete global identifications of spacelike stretched $AdS_3$. They are analogues for the case of warped $AdS_3$ of the BTZ black hole in $AdS_3$. In fact, when $\nu = 1$ the metric becomes the BTZ metric in a rotating frame. We note that slices of constant $t$ and $\theta$ are both spacelike for $\tilde{r} > r_+$. However, the metric is everywhere Lorentzian.

The mass and angular momentum are related to $r_+$ and $r_-$. There is  a continuous spectrum of black holes all the way to $r_+ = r_- = 0$, where we find warped $AdS_3$ in Poincar\'e coordinates with $\tau$ identified. Following the analogy with BTZ in $AdS_3$, we would expect that lowering the energy below the black hole continuum we should find a mass gap, with global stretched $AdS_3$ being the ground state. This is in fact not the case, and it can be shown that there are no values of $r_{\pm}$ for which the metric is both real and has a global $SL(2,\mathbb{R}) \times U(1)$ isometry. Thus, the vacuum (unquotiented) space is not part of the family of spacetimes \eqref{ng}. This is in agreement with the fact that global warped $AdS_3$    has no Killing vectors with compact orbits, a property that the metrics \eqref{ng} clearly do not share\footnote{We are grateful to A. Strominger for making this point.}.

It has been shown \cite{Bouchareb:2007yx} that the black holes \eqref{ng} obey the first law of thermodynamics, once one employs the Chern-Simons corrected entropy formula \cite{Kraus:2005vz,Solodukhin:2005ah,Tachikawa:2006sz}. In \cite{Anninos:2008fx} this entropy formula has been suggestively rewritten as the entropy of a thermal state in a two-dimensional CFT with unequal left-and right-moving central charges
\be
c_R = \frac{(5 \nu^2 +3)\ell}{\nu (\nu^2+3) G} \;, \;\;\;\;\; c_L = \frac{4\nu \ell}{ (\nu^2+3)G} \label{cleft}
\ee

\subsection{Asymptotic behavior}

In order to answer questions about boundary conditions, we need to understand where the boundary circle lies in the geometries of interest, since the conserved charges are constructed as integrals over this circle.

Let us start with the easy case, which are the warped black holes of the previous section. The  boundary circle consists of two disconnected pieces, one at $\tr \r \infty$ and one at $\tr \r - \infty$, and each is parameterized by $\th$. Asymptotically, the warped black hole metric can be written as
      \be
      \frac{ds^2}{\ell^2} = \frac{3(\nu^2-1) \tr^2 d\theta^2}{4}  +  \frac{d\tr^2}{(\nu^2+3)\tr^2}  + dt^2 + 2\nu \tr
      dtd\theta +  h_{\mu\nu} dx^\mu dx^\nu \label{ngas}
      \ee
      where the `perturbation' $h_{\mu\nu}$ falls off at least one power of $\tr$ faster than the background \cite{Compere:2008cv}. The boundary conditions defined near this boundary are reasonably well understood.

Note that the asymptotic form of \eqref{ng} and \eqref{sqads3} are the same, given that one identifies\footnote{We define $r_0 = \frac{\sqrt{r_+r_-(\nu^2+3)}}{2 \nu}$.}
\be
\tau \leftrightarrow \frac{(\nu^2+3)}{2} \theta \;, \;\;\;\; r \leftrightarrow \tr \;, \;\;\;\; x \leftrightarrow  \frac{(\nu^2+3)}{2\nu} ( t - \nu r_0 \,\th)
\ee
Nevertheless, for global warped $AdS_3$ both $\tau$ and $x$  are noncompact, so this identification only holds locally at asymptotic infinity. Moreover, it is not quite clear where the boundary circle lies in these coordinates. If we consider a surface of constant $\tau$ - which is always spacelike and always intersects the boundary, we find that the induced metric is
\be
ds^2_{\tau} = a^2 dx^2 + d \sigma^2 \;, \;\;\;\;\; \sigma= \sinh^{-1} r
\ee
Thus, the boundary circle consists of four pieces:
\be
r \r \pm \infty\,, \,x \; \mbox{finite}\;, \;\;\;\;\;\;\; x \r \pm \infty \,, \,r \;\mbox{finite}
\ee
This is the same conclusion that one reaches when studying the boundary of $AdS_3$ in fibered coordinates (see appendix A). Quotienting along various Killing vectors \cite{Anninos:2008fx} gives the black holes \eqref{ng}. These quotients act on the boundary and split it into two disconnected circles at $r \r \pm \infty$.

\section{Preliminaries II: first order perturbation theory}\label{pert}

Having discussed the background geometry and its asymptotic structure, we now delve into the linearized equations of motion. We also define the notion of energy, for both propagating and pure large gauge solutions, that we will use to test stability.

\subsection{Linearized perturbations around warped $AdS_3$}

Our goal is to study linearized perturbations around the background \eqref{sqads3}, and find out whether they can destabilize the spacetime. The linearized equations of motion for a metric perturbation $h_{\mu\nu}$ read
\be
R^{(1)}_{\mu\nu} -( \frac{1}{2}  R +\frac{1}{\ell^2})  h_{\mu\nu} - \frac{\ell}{3 \nu} C_{\mu\nu}^{(1)}=0 \label{lineom}
\ee
where
\be
R^{(1)}_{\mu\nu} = \frac{1}{2} (\nabla^{\l} \nabla_{\mu}h_{\l\nu}+\nabla^{\l} \nabla_{\nu}h_{\l\mu}-\nabla^{\l} \nabla_{\l}h_{\mu\nu}-\nabla_{\mu} \nabla_{\nu}h^{\mu}_{\mu}) \non
\ee
\be
C_{\mu\nu}^{(1)} = \e_{\mu}{}^{\a\b} \nabla_{\a} (R_{\b \nu}^{(1)} - \frac{1}{4} h_{\b\nu} R) - \e_{\mu}{}^{\a\b} \delta \Gamma_{\a\nu}^{\l} (R_{\b \l} - \frac{1}{4} g_{\b\l} R)
\ee
\be
\delta \Gamma_{\mu\nu}^{\l} = \half(\nabla_{\mu} h^{\l}{}_{\nu}+ \nabla_{\nu}h^{\l}{}_{\mu} -\nabla^{\l} h_{\mu\nu})
\ee
All derivatives are taken with respect to the background metric \eqref{sqads3}.

When studying perturbations around $AdS_3$, the fact that the background is maximally symmetric drastically simplifies the third order linearized equations of motion. It can then be shown \cite{Li:2008dq} that if one chooses to work in harmonic gauge
\be
\nabla_{\mu} h^{\mu\nu} =0 \label{harmg}
\ee
the equations of motion \eqref{lineom} take the form $D_{M} D_L D_R h_{\mu\nu} =0$, where $D_I$ are three commuting linear differential operators. Of the three distinct solutions that one gets for generic $\mu$ away from the chiral point, only one describes the propagating massive graviton, while the other two are pure (large) gauge \cite{Li:2008dq}. It is apparent that the  condition \eqref{harmg} did not completely fix the gauge redundancy in the problem.

Due to the fewer symmetries of the warped $AdS_3$ background, we were unable to bring the equations of motion to an analogously simple form. Nevertheless, we may expect the third order equations of motion to split at least into a first order piece and a second order one - the reason being that TMG does describe one propagating mode, which should obey a second order wave-like equation. The most elegant option not readily feasible, we have decided to attack these equations on two fronts:
\bi
\item fix the gauge completely and solve the linearized equations of motion. In this way, we are sure to only be describing the propagating mode.
\ei
In agreement with our expectations, the equations of motion decouple\footnote{We have assumed though that we can concentrate solely on separable momentum eigenstates.}. While the equation we obtain is in principle tractable, for the purposes of this article it is only the asymptotic behavior of the solution that is relevant, so we leave its full analysis  for subsequent work.
\bi
\item use the $SL(2,\mathbb{R})_R \times U(1)_L$ isometry of the background to classify perturbations. As is well known, linearized solutions to the equations of motion must fall into representations of the isometry group of the background. The most commonly encountered representation of the above isometry group is the highest weight one. We thus consider a basis of perturbations $\psi_{\mu\nu}$ that are eigenfunctions of $U(1)_L$ and belong to a highest weight representation of $SL(2,\mathbb{R})_R$
\be
J_0 \psi_{\mu\nu}= k \psi_{\mu\nu} \;, \;\;\;\;\; \tilde{L}_0 \psi_{\mu\nu} = \omega \psi_{\mu\nu}\;, \;\;\;\;\;\tilde{L}_1 \psi_{\mu\nu}=0 \label{highestwt}
\ee
\ei
While it is not true that all solutions to the equations of motion can be written as a superposition of  $SL(2,\mathbb{R})_R$ highest weight states and their descendants, such perturbations are a physically relevant subclass of solutions, especially from the point of view of the AdS/CFT correspondence. Even though our analysis is not exhaustive, it still proves sufficient for the purposes of this article, as will soon become clear.
\medskip
The solutions of physical interest can be split into two types:
\bi
\item {\it propagating}, if the metric perturbation cannot be written as $\psi_{\mu\nu} = \mathcal{L}_{\xi} \,\bar{g}_{\mu\nu}$ for any diffeomorphism $\xi^{\mu}$. Here  $\bar{g}_{\mu\nu}$ is the background metric \eqref{sqads3}.
\item {\it pure large gauge}, if  $\psi_{\mu\nu} = \mathcal{L}_{\xi}\, \bar{g}_{\mu\nu}$ for some diffeomorphism $\xi^{\mu}$ that does not vanish `sufficiently rapidly' at asymptotic infinity.
\ei

The propagating modes are easy to find: one first chooses a gauge such that {\it all} the gauge freedom in choosing the metric components is fixed, and then proceeds to look for solutions to \eqref{lineom} which have a wavelike behavior. In appendix B we describe such a gauge fixing for the case in which the gravitational waves have a nontrivial dependence on the coordinate $x$. An appropriate gauge-fixing for the $x$-independent case has been described in \cite{Kim:2009xx}.

Since we will concentrate our attention on highest weight solutions, we would like to find a gauge which preserves the highest weight property. This basically requires that we gauge-fix only using diffeomorphisms that commute with $\tL_1$ and have appropriate weights under $\tL_0$ and $J_0$.

Pure large gauge perturbations automatically satisfy the linearized equations of motion. Whether they are physically relevant is determined entirely by the choice of boundary conditions, as will be reviewed in the next section. The pure large gauge modes are by definition those which generically carry nontrivial conserved charges as measured at infinity, and thus are physically relevant. For example, in pure Einstein gravity in $AdS_3$, propagating modes do not exist and it is precisely the pure large gauge modes that correspond - upon quantization - to states in the dual CFT.  A general procedure for computing the conserved charges, finding a set of consistent boundary conditions and determining which are the `large' gauge transformations is presented below.

\subsection{Consistent boundary conditions and the asymptotic symmetry group}\label{asgbc}

The quest for consistent boundary conditions for the metric perturbations in a given background spacetime proceeds in several steps:
\begin{enumerate}
\item locate the boundary of the background spacetime
\item impose boundary conditions on the metric fluctuations at asymptotic infinity
\item find all the diffeomorphisms that preserve the boundary conditions
\item show that the boundary conditions are consistent, which requires computing all the charges associated with the asymptotic symmetry generators and showing that they are {\it conserved, finite} and {\it integrable}. If infinities are found, one may need to impose additional boundary conditions, or altogether change the ones that were originally proposed. The asymptotic symmetries are defined to be those allowed diffeomorphisms which have nonzero conserved charges on a generic allowed background. \item find the Dirac bracket algebra of the asymptotic generators, which form the asymptotic symmetry group (ASG) \cite{Brown:1986nw}, together with its eventual central extension.
\end{enumerate}

The conserved charges $Q_{\xi}$ associated with the asymptotic symmetry generators $\xi^{\mu}$ can be computed in a variety of ways. A particularly nice formalism has been developed in \cite{Barnich:2001jy}. If $\bar{g}_{\mu\nu}$ denotes the background metric and $h_{\mu\nu}$ a perturbation satisfying the boundary conditions, then the conserved charges $Q_{\xi} [h,\bar{g}]$ are constructed as surface integrals over the spacelike boundary $\p \Sigma$ of the ($n$-dimensional) spacetime
\be
Q_{\xi} [h,\bar{g}] = \int_{\p \Sigma} K^{(n-2)}_{\xi}[h,\bar{g}] \label{bbchl}
\ee
where $K^{(n-2)}$ is a particular $n-2$ form constructed from  the linearized equations of motion for $h_{\mu\nu}$. The explicit expression for $K^{(1)}$ for TMG is given in \cite{Compere:2008cv}.

The above expression for the charges is valid for finite (as opposed to infinitesimal) $h_{\mu\nu}$ when a certain property called asymptotic linearity holds, which takes the form
\be
Q_{\xi} [h,\bar{g}] = Q_{\xi} [h,\bar{g} + \delta g] \;, \;\;\; \forall \xi \in \mbox{ASG}
\ee
where $\d g_{\mu\nu}$ is any perturbation of the background metric consistent with the boundary conditions. This is because the expression \eqref{bbchl} was derived for linearized perturbations $\d h_{\mu\nu}$ around a given background, and one needs to integrate over a path in phase space in order to find the charges associated with finite departures from the background metric. Thus, the general expression for the charges is
\be
Q_{\xi} [h,\bar{g}] = \int_{\gamma} D \d g\int_{\p \Sigma} K^{(n-2)}_{\xi}[\d g,g(\gamma)] \label{bbchgen}
\ee
where $\gamma$ is a path in phase space which connects $\bar{g}_{\mu\nu}$ to $\bar{g}_{\mu\nu}+h_{\mu\nu}$. The above integral only makes sense if it is independent of the path $\gamma$, which reduces to the requirement of charge integrability
\be
Q_{\xi} [\d h_1,g+\d h_2] - Q_{\xi} [\d h_2,g+\d h_1]-Q_{\xi} [\d h_1+\d h_2,g] =0
\ee
for {\it any} background metric $g_{\mu\nu}$ allowed by the boundary conditions.

Integrability needs to be checked for warped $AdS_3$, as the theory is not asymptotically linear in this background. Besides integrability, one also needs to check finiteness of the generators \eqref{bbchgen} and conservation\footnote{In fact, conservation of the charges $Q_{\xi}$ should hold by construction; however for relaxed boundary conditions one must check conservation by hand.}.

\subsection{The energy-momentum pseudo-tensor}

In order to address the question of stability, we need to compute the energy of the various gravitational perturbations \cite{Abbott:1981ff,Deser:2002rt,Deser:2002jk,Deser:2003vh} and find its sign. Using the formalism from the previous section, the energy in question is just the charge associated to the relevant Killing vector ($\p_{\tau}$ in this case) of the back-reacted gravity solution.

In this section we follow the discussion in \cite{Maloney:2009ck}. If we let $\xi^{\mu}\partial_\mu = \p_{\tau}$ and $h_{\mu\nu}$ be some perturbation of the background metric $\bar{g}_{\mu\nu}$, then it can be shown that the expression for the energy is
\be
Q_{\xi}[h,\bar{g}]=  \frac{1}{16 \pi G} \int_{\Sigma} \star (\xi^{\mu} E_{\mu\nu}^{(2)} [h^{(1)}] d x^{\nu} )
\ee
where $E_{\mu\nu}^{(2)}$ are the TMG equations of motion at second order in perturbation theory, evaluated on a solution $h^{(1)}_{\mu\nu}$ of the linearized equations of motion. $\Sigma$ is a spatial slice at constant $\tau$ and $\star$ denotes the Hodge star operation. The quantity $-E_{\mu\nu}^{(2)}(h^{(1)})$ is sometimes also called the energy-momentum pseudo-tensor, because it sources the linearized equations of motion for the second order metric perturbation. It reads
\be
E_{\mu\nu}^{(2)} = G_{\mu\nu}^{(2)} + \frac{\ell}{3\nu} C_{\mu\nu}^{(2)}
\ee
where $G_{\mu\nu}^{(2)} ,C_{\mu\nu}^{(2)}$ are the Einstein and the Cotton tensor, respectively, evaluated to second order in the perturbation $h^{(1)}_{\mu\nu}$.
Since we are building plane waves along $x$, what we will actually be computing is the {\it energy density} of a gravitational wave in the $x$-direction, which is given by
\be
\mathcal{E}_P = \frac{1}{16 \pi G} \int dr \sqrt{-g} \,g^{\tau \mu} E_{\mu\nu}^{(2)} \xi^{\nu} \label{engd}
\ee
and try to establish its sign.

\section{The massive gravitons}\label{massgrav}

We find explicit highest weight solutions and the asymptotic structure for all solutions to the linearized equations of motion. The energy density of the highest weight solutions is found to be negative. However, we discard the modes by imposing boundary conditions.

\subsection{Highest weight solutions}

We consider the following Ansatz for the metric perturbation
\be
\psi_{\mu\nu} (\tau,r,x) = f_6(r) e^{i(k x-\om \tau)} \, \left(\begin{array}{ccc} f_1(r) & f_2(r) & f_3(r) \\ f_2(r) & f_4(r) & f_5(r) \\ f_3(r) & f_5(r) & C_6 \end{array} \right) \label{genform}
\ee
If we first solve the highest weight condition, we find that the functional form of the solution is completely fixed, up
to six constants $C_i$, $i=\{1, \ldots, 6 \}$, subject to rescaling by an overall factor. The function $f_6(r)$ takes the form
\be
f_6(r) = e^{k \,  \tan^{-1} r} (1+r^2)^{- \frac{\om}{2}}
\ee
while the ratios of the metric components take a relatively simple form
\bea
f_1(r) & = & C_5 + r\, C_4 + r^2 \,C_3 \non \\
f_2(r) & = & \frac{i}{2(1+r^2)} \left( C_4-2C_1 + 2 r \,(C_3-C_2-C_5)  - r^2\, C_4  \right) \non \\
f_3(r)&=& C_1 + r\, C_2 \non \\
f_4(r) &=& \frac{1}{(1+r^2)^2} (2 C_2 -C_3-C_6 + r\,(C_4 -2 C_1)  - r^2 \,C_5)\non \\
f_5(r) & = & \frac{i}{1+r^2} (C_2 - C_6 - r\, C_1) \label{gensolhw}
\eea
Next, we  gauge-fix in such a way that the highest weight condition is preserved. It turns out that as long as
\be
k(k^2+a^4) \neq 0 \label{gauging}
\ee
then one can always set $C_6=f_3(r)=f_5(r)=0$ (actually $f_5=0$ follows from the previous two conditions). This amounts to setting three of the constants ($C_1=C_2=C_6$) to zero. Next, one plugs the gauge-fixed highest weight perturbation into a subset of the linearized equations of motion, which can be written as
\be
A v =0\;, \;\;\;\; A \in \mathcal{M}_{3\times 3} \;, \;\;\;\; v = \left( \begin{array}{c} C_3 \\ C_4 \\ C_5 \end{array} \right)
\ee
The determinant of $A$ is
\be
\det A = k^2 (a^4 + k^2)(k^2 + (\om-2)^2)(a^4 - k^2 + a^2 (k^2-1 - \om + \om^2))
\ee
Thus, if $\det A \neq 0$, the only solution is $C_3=C_4=C_5=0$ and  we conclude that all highest weight modes of this form are pure gauge. If $\det A =0$ we get a non-pure gauge mode, obtained when $v$ is in the kernel of $A$. The constants $C_{3,4}$ are then determined in terms of $C_5$ via the equations of motion.

Since we are looking for square integrable solutions, $\psi_{\mu\nu}$, and the range of $x$ is infinite, $k$ must be real. For the classical analysis we are performing, we should also take $\om$ to be real. Thus we obtain a propagating mode only if
\be
\om = \omega_{\pm} \equiv \frac{1}{2} \pm \sqrt{ k^2\left(\frac{1}{a^2}-1\right) + \frac{5}{4}-a^2} \label{omsol}
\ee
For convenience, let us also define
\be
\omega_{1,2} \equiv \frac{1}{2} \pm \sqrt{ \frac{5-4 a^2}{4}}
\ee
Given that  $\omega, k \in \mathbb{R}$, we can distinguish several cases
\bi
\item $a < 1 $.  All values of $k$ are allowed, while
\be
\omega \in (- \infty, \om_2) \cup (\om_1, \infty)
\ee
In this particular squashing parameter range, we have $1< \om_1 < (1+\sqrt{5})/2$ and $(1-\sqrt{5})/2 < \om_2 < 0$.

\item $1< a < \frac{\sqrt{5}}{2}$. In order to have $\om \in \mathbb{R}$, $|k|$ must be bounded above as

\be
k^2  \leq \frac{5 - 4 a^2}{4 (a^2-1)} \, a^2 \;, \;\;\; \mbox{while} \;\;\;\; \om_2 \leq \om \leq \om_1
\ee
In this parameter range, we have $\half < \om_1 <1$ and $0< \om_2 < \half$.

\item $ a > \frac{\sqrt{5}}{2}$. Then there is no highest weight propagating solution to the linearized TMG equations of motion in this background.

\ei
In summary, a propagating highest weight mode takes the form
\be
\psi_{\mu\nu} (\tau,r,x) = \frac{1}{(1+r^2)^{\frac{\om}{2}}} e^{i(k x - \om \tau)+ k \tan^{-1} r } \, \left(\begin{array}{ccc} f_1(r) & f_2(r) & 0 \\ f_2(r) & f_4(r) & 0 \\ 0 & 0 & 0\end{array} \right) \label{propmode}
\ee
with $\om$ given by \eqref{omsol}, $f_{i}(r)$ by \eqref{gensolhw} with $C_1=C_2=C_6=0$ and $C_{3,4}$ given in terms of $C_5$ in the appendix. Moreover, $k$ is subject to the restrictions mentioned above.

\subsection{Energy}

Our next task is to compute the energy of these modes using \eqref{engd}. Since the physical metric perturbation must be real, we take
\be
h_{\mu\nu} = \a\, \psi_{\mu\nu} + \a^*( \psi_{\mu\nu})^*
\ee
Also, given that our wave solutions are $\partial_x$ eigenfunctions and the energy is obtained by integrating over a whole spatial slice at constant $\tau$, the energy of a single $k$-mode will diverge. Thus, it is more appropriate to consider the energy density of the modes per unit length in the $x$-coordinate. Finite energy configurations are then obtained by generating linear combinations of the $k$-modes with compact support in $x$.

The expression for the energy density is analytically tractable but extremely lengthy and unilluminating. Its salient feature is that it can be written as a rather simple integral:
\be
\mathcal{E}_P =  |\a C_5|^2  \int dr \frac{e^{2 k \, \tan^{-1} \, r}}{ (1+r^2)^{\om+4} }\sum_{n=0}^{8} r^n  \,b_n (k, a) \equiv \sum_{n=0}^{8} I_n(k,a)  \,b_n (k, a)
\ee
where we have defined
\be
I_n(k,a) \equiv \int_{-\infty}^{\infty} dr \frac{r^n \, e^{2 k \, \tan^{-1} \, r}}{ (1+r^2)^{\om+4}}\;, \;\;\;\;\;\; \om = \om (k,a)
\ee
One finds that the above integrals are non-divergent if $Re [\om] > \half$ and $n \leq 8$. Thus we should only restrict ourselves to the upper branch ($\om_+$) of solutions \eqref{omsol}, given that the lower branch ($\om_-$) always has divergent energy density\footnote{The `superradiant' modes with complex $\om_{\pm}$ have $Re[\om]=\half$, so they carry infinite energy density and thus must be discarded.}. For $n\leq 8$, the integrals $I_n$ obey a recursion relation
\be
(\Omega - n+1 ) I_n = 2 k I_{n-1} + (n-1) I_{n-2} \;, \;\;\;\;\;\; \Omega = 2 \om + 6
\ee
so we can rewrite the whole expression for the energy in terms of $I_0(k,a)$. It turns out that for the entire allowed range of $k$ $I_0(k,a)$ is real and positive, whereas the coefficient multiplying it is always negative. We thus conclude that
\be
\mathcal{E}_P <0 \;, \;\;\; \forall \, a \in (0,2)\;\; \& \;\; \forall \, k \; \mbox{allowed.}
\ee

\subsection{The asymptotics save the day}

At first sight, the fact that the energy of the propagating highest weight modes is always negative may sound discouraging as far as the stability of warped $AdS_3$ is concerned. Nevertheless, an important point is that a theory is not only defined by its Lagrangian, but also by boundary conditions to be imposed on the fields.

Consider the expression \eqref{propmode} for the highest weight modes. The asymptotic behavior of the metric perturbation is\footnote{Note that when $\om_+-\om_- \in \mathbb{Z}$ then logarithmic asymptotic behavior is also allowed. This occurs in stretched $AdS_3$ only for $a< \frac{\sqrt{5}}{2}$ and at isolated values of $k$ in squashed $AdS_3$.}
\be
h_{\tau\tau} \sim r^{2-\om} \;, \;\;\;\; h_{\tau x} \sim r^{1-\om} \;, \;\;\;\;h_{\tau r} \sim r^{-\om}\;,\non
\ee
\be
h_{xx} \sim r^{-\om} \;, \;\;\;\; h_{r r} \sim r^{-2-\om} \;, \;\;\;\;h_{r x} \sim r^{-1-\om} \label{falloffhw}
\ee
Note that as \mbox{$r \rightarrow \pm \infty$} the perturbation roughly  falls off by a factor of $r^{-\om}$ faster than the components of the background metric. A set of boundary conditions for stretched $AdS_3$ proposed in \cite{Compere:2008cv,Compere:2007in}, and further elaborated in section \ref{boundaryconditions}, require that perturbations fall off at least by one power of $r$ faster than the background metric. Thus we are instructed to only include those highest weight modes which have
\be
\om (k, a) \geq 1
\ee
in the spectrum.

For stretched $AdS_3$,  $a > 1$ and $\om (k, a)  < 1$.  Thus we can conclude that {\it all the highest weight, negative energy propagating modes are excluded from the spectrum of TMG in stretched $AdS_3$ backgrounds}. We will see in the following section that we can in fact exclude {\it all} propagating modes from the spectrum of TMG if $a>1$.

Notice that the presence of even a single negative energy mode in stretched $AdS_3$  renders the theory unstable. The reason is that the $a>1$ theory contains black holes \cite{Anninos:2008fx}, which have {\it positive} energy in our conventions. It has been quite customary in the context of TMG \cite{Deser:1981wh,Deser:1982vy} to reverse the sign of Newton's constant $G$, which amounts to reversing the sign of the energy. In asymptotically flat space, the only allowed configurations are negative energy massive gravitons, and switching the sign of $G$ seems to render them harmless for the stability of TMG about flat space. The sign of Newton's constant is fixed by requiring that the black holes have positive energy, and no further discussion is possible.

In the case of squashed $AdS_3$, $a<1$ and all modes have $\om(k,a) > 1$. Thus they are all allowed by the boundary conditions \eqref{compbc} so long as we take wave packets with compact support in $x$. Nevertheless, it is unclear whether these boundary conditions make sense in the case of squashed $AdS_3$.

Finally, for $a=1$ - which is just $AdS_3$ - the negative energy modes have $\om=1$ and are thus allowed by the boundary conditions, both \eqref{compbc} and Brown-Henneaux\footnote{At $a = 1$ the boundary conditions \eqref{compbc} are more restrictive than Brown-Henneaux boundary conditions, as they only allow half of the usual $AdS_3$ asymptotic symmetry group.}. Nevertheless, we already knew that the propagating mode would have negative energy - since the limiting $AdS_3$ case must have $\mu \ell = 3$, so it is away from the chiral point.

\subsection{General propagating solutions}

We will now decouple the linearized equations for the propagating mode. Again, we assume that the perturbation of interest is an eigenmode of energy and $x$-momentum, and thus it can be written as $h_{\mu\nu} (\tau, r,x) = e^{i (k x-\om \tau) } \tilde{h}_{\mu\nu}(r)$. The $k=0$ case has already been found in \cite{Kim:2009xx} and will be discussed in the next subsection. Whenever $k \neq 0$, we can safely impose the gauge
\be
h_{\mu x} =0
\ee
as shown in appendix B. We therefore consider the metric Ansatz
\be
h_{\mu\nu} (\tau, r,x) =  e^{i (k x-\om \tau )}  \left(\begin{array}{ccc} - (1+r^2) g_1(r) & g_2(r) & 0 \\ g_2(r) &  (1+r^2)^{-1} g_3(r) & 0 \\ 0 & 0 & 0  \end{array}\right)
\ee
The coupled system of equations \eqref{lineom} decouples as follows
\be
g_3{}''(r) + A(r)\, g_3{}'(r) + B(r)\, g_3 =0 \label{nastyeom}
\ee
where
\be
A(r) = \frac{P_5(r)}{(1+r^2) P_4(r)} \;, \;\;\;\;\; B(r) = \frac{P_6(r)}{(1+r^2)^2 P_4(r)}
\ee
and $P_n(r)$ are $n$th degree polynomials in $r$, whose coefficients depend on $k$, $a$ and $\omega$. The expressions for $P_{4,5,6}(r)$ are given in appendix C. The existence and regularity of the solutions to \eqref{nastyeom} can be analyzed using Frobenius' method. For more details we refer the reader to appendix D. All we need from \eqref{nastyeom} is the asymptotic behavior of the solutions. Following appendix D, we know that as $r \r \pm \infty$ the solutions behave as
\be
g_3(r) = \frac{1}{r^{\alpha}} \sum_{s=0}^{\infty} \frac{a_s}{r^s}
\ee
where $\a$ is a solution to the indicial equation
\be
\a (\a+1) - a_0 \a + b_0 =0
\ee
with
\be
a_0 = \lim_{r \r \infty}  r A(r) =2 \;, \;\;\;\;\; b_0 = \lim_{r\r \infty} r^2 B(r) = \frac{(a^2-1){k^2+a^2}}{a^2}
\ee
It is easy to see that the solutions are simply $\a_{\pm} = \om_{\pm}$ defined in \eqref{omsol}. Consequently, near $r \r \pm \infty$, the solution behaves as
\be
g_{3} (r) \sim r^{-\om_{\pm}}
\ee
The remaining metric components have a similar asymptotic behavior $g_{1,2}(r) \sim r^{-\om_{\pm}}$. Notice that we recover precisely the asymptotic behavior \eqref{falloffhw} of the highest weight modes. There is one difference though, in that highest weight modes were obliged to have $\om = \om_{\pm} (k, a)$, whereas no such relation is necessary in the case of general propagating modes.

In fact, if we  consider an arbitrary but decoupled time dependence for the mode, i.e. $h_{\mu\nu}(\tau,r,x) = f(\tau) e^{i k x} \tilde{h}_{\mu\nu}(r)$, the equations of motion imply the same asymptotic falloff for $r$ as they did for the energy eigenstates. This agrees with the expectation that we can construct generic solution by superimposing various highest weight modes and their descendants, and possibly modes that belong to different $SL(2,\mathbb{R})$ representations\footnote{In the analysis of \cite{Balasubramanian:1998sn} it was noted that $\omega$ obeyed a quantization condition, $\om = \om_+ + n, \; n \in \mathbb{Z^+}$. This quantization condition stemmed from the requirement that the solution be well-behaved near the origin of $AdS_3$. Due to the complexity of our equations, we have been unable to obtain a similar quantization condition, although it is likely that it exists. }.

In conclusion, we find that the most general propagating solution to the linearized equations of motion of TMG has the same asymptotic behavior as the highest weight modes do. The discussion in section 4.3 still applies and, for stretched $AdS_3$, we can invoke boundary conditions which exclude {\it all} propagating modes from the spectrum. The remaining pure gauge modes form the subject of the next section.

\subsection{Solutions with $k=0$}

Before we move on to the pure gauge modes, let us make a few comments on propagating solutions with $k=0$. As emphasized in appendix B and elsewhere, our gauge-fixing condition does not apply in this case.

The appropriate gauge-fixing condition and equations of motion were written down in \cite{Kim:2009xx}, which studied the problem by dimensionally reducing it to propagation in $AdS_2$. The authors found that the propagating mode obeys an equation of the form
\be
\Box_2 \phi - m^2 \phi =0 \;, \;\;\;\;\; m^2 = - \frac{3(\nu^2-1)}{\ell^2} \label{scscor}
\ee
It is not hard to show that the asymptotic falloff of the solution to this equation is the same as our \eqref{falloffhw}, if we set $k=0$. Thus, propagating modes with $k=0$ are allowed or disallowed by the boundary conditions just as their $k \neq 0$ counterparts. This is to be expected, as different ways of gauge-fixing should not affect the allowed spectrum of the theory\footnote{In the case of the propagating modes, one can check whether the falloff of the gauge-invariant quantities respects the boundary conditions.}.
\medskip
One further check that the equation \eqref{scscor} of \cite{Kim:2009xx} and our results indeed agree, is to compare the conformal weight and mass  of the $k=0$ graviton. Consider a scalar field of mass $m$ which propagates in spacelike warped $AdS_3$. We look for a highest weight solution of the form
\be
\Phi(\tau,r,x) = e^{i(k x - \om \tau)} \phi(r)
\ee
The conformal weight of the scalar is determined in terms of the mass $m$ as
\be
\om = \frac{1}{2} + \sqrt{ k^2\left(\frac{1}{a^2}-1\right) + \frac{1}{4} + L^2 m^2} \;, \;\;\;\; L= \frac{\ell}{\sqrt{\nu^2+3}}
\ee
Comparing the above result with $\om_+$, we note that the graviton behaves as a scalar field of mass
\be
m^2= - \frac{3(\nu^2-1)}{\ell^2}\label{mass}
\ee
This coincides exactly with the result of \cite{Kim:2009xx} for the $k=0$ case.

\section{Boundary gravitons}\label{boundgrav}

Here we propose a set of boundary conditions for global stretched  $AdS_3$, which are a slightly modified version of those put forth in \cite{cdwip} for the asymptotic black hole spacetime \eqref{ngas}.

\subsection{Boundary conditions for stretched $AdS_3$\label{boundaryconditions}}

The boundary conditions we use for the asymptotic metric (rescaled by $(\nu^2+3) l^{-2}$) at large $r$ are
\be
g_{\tau\tau} = (a^2-1) r^2 + r\, h_{\tau\tau} + \O(r^0) \;, \;\;\;\;\;  g_{\tau r} = r^{-1} h_{\tau r} + \O(r^{-2}) \non
\ee
\be
g_{\tau x} = a r + h_{\tau x} + \O(r^{-1}) \;, \;\;\;\;\; g_{xx}= a^2  + r^{-1} h_{xx} + \O(r^{-2}) \;, \non
\ee
\be
g_{xr} = \O(r^{-2}) \;, \;\;\;\;\; g_{rr} = r^{-2} + r^{-3} h_{rr} + \O(r^{-4})  \label{compbc}
\ee
The perturbations $h_{\mu\nu} (\tau,x)$ of the background metric would generically yield nontrivial conserved charges, while the terms written as $\O(r^{n})$ do not contribute to the conserved quantities.

These boundary conditions were developed in \cite{Compere:2008cv,Compere:2007in} for the black hole metrics, where $\tau$ is identified and the boundary lies at large $r$ only. As noted in section \ref{wbtz}, global stretched $AdS_3$ is not part of the black hole phase space, and the boundary has a piece that lies at $r$ finite and  $x \r \pm \infty$. Therefore, the boundary conditions listed above must be supplemented by conditions on the falloff of the metric components at large $x$. We have not studied precisely what these restrictions look like, but we always consider wave packets which die off sufficiently fast\footnote{For example, a falloff as $|x|^{-2}$ at large $|x|$ seems more than sufficient to ensure the finiteness of the charges.} as $x \r \pm \infty$.

Following our checklist from section \ref{asgbc}, we need to make sure that the above boundary conditions yield charges which are finite, integrable and conserved. A first thing to note is that TMG with the above boundary conditions is not asymptotically linear. Thus, integrability, finiteness and conservation of the charges for all {\it finite} $h_{\mu\nu}$ that obey the asymptotic equations of motion must all be considered. This has been rigorously done in \cite{cdwip} for asymptotically warped black hole geometries. As we show in appendix E, for fast enough asymptotic falloff at large $|x|$ the charges do not gain ay additional contributions from the integrals along $x$ or $r$, so consistency of the boundary conditions for global stretched $AdS_3$ follows from consistency of the very related boundary conditions for the black holes.

\subsection{Pure large gauge modes}

In this section we will be exclusively working in the warped black hole coordinate system \eqref{ng}. Given that the boundary conditions \eqref{compbc} have excluded all propagating modes from stretched $AdS_3$, all the remaining physical excitations in our theory must correspond to pure large gauge modes: diffeomorphisms that do not fall off fast enough near the boundary at $\tilde{r} \r \infty$. In \cite{Compere:2008cv}, they were shown to take the form
\be
\xi^{\th} = f(\th) + \O(\frac{1}{\tr^{2}}) \;, \;\;\;\;\; \xi^{\tr} = - \tr f'(\th) + \O(1) \;, \;\;\;\;\; \xi^{t} = g(\th) + \O(\frac{1}{\tr})
\ee
As reviewed in section \ref{asgbc}, to each non-trivial large diffeomorphism there corresponds a generator of the asymptotic symmetry group. By expanding $f(\th),g(\th)$ in Fourier modes, \cite{Compere:2008cv} found the ASG to consist of one copy of the Virasoro algebra and a $U(1)$ Ka\v{c}-Moody algebra. The Virasoro acquires a central extension, with positive central charge
\be
c_R  = \frac{(5\nu^2+3) \ell}{\nu(\nu^2+3)G} >0
\ee
which precisely confirms (half of) the conjecture in \cite{Anninos:2008fx} and moreover ensures that all energies are positive. We thus conclude that stretched $AdS_3$ with the boundary conditions \eqref{compbc} is {\it stable}.

For $\nu<1$, $\th$ cannot be identified in the coordinates \eqref{ng}, as the resulting spacetime would have CTCs. We can thus no longer use the boundary conditions of \cite{Compere:2008cv}. The boundary conditions \eqref{compbc} could in principle still hold, since our coordinate $\tau$ is noncompact. Thus, if \eqref{compbc} are extendable to a full set of consistent boundary conditions for squashed $AdS_3$, then we conclude that the latter spacetime is \textit{unstable}, since the negative energy propagating modes are not excluded. Whether such an extension is possible for squashed $AdS_3$
is very unclear.

\section{Summary and open questions}\label{discussion}

In this note, we have addressed the issue of the stability of spacelike warped $AdS_3$. We have found explicit propagating massive gravitons living in the highest weight representation of the $SL(2,\mathbb{R})_R$ isometry group and the asymptotic falloff of all propagating momentum eigenstates.

Our highest weight solutions obey an equation of the form
\be
\left[ \frac{1}{2}\left(L_+ L_- + L_- L_+\right) - L^2_0 + \frac{(1-a^2)}{a^2}\partial^2_u \right] \psi_{\mu\nu}= -(1-a^2) \psi_{\mu\nu}
\ee
We have not managed to recover such an elegant equation in terms of the quadratic Casimir from the general linearized equations of motion. The only exception is for $k=0$, where it was obtained by \cite{Kim:2009xx}. We  suspect, however, that under the appropriate gauge choice and field redefinition this is possible for all values of $k$. The fact that we have obtained a decoupled second order wave equation for a single component of the perturbation is evidence in this direction.

The explicit solutions we have obtained have negative energy density, however for $a >1$ they do not obey the set of boundary conditions discussed in section \ref{boundgrav}. Imposing those boundary conditions leads us to discard all propagating solutions from the physical spectrum about spacelike stretched $AdS_3$. We take this to be strong evidence that TMG has a stable set of vacua for a much larger range of the Chern-Simons coefficient, i.e. $\mu\ell > 3$, than simply the chiral point $\mu\ell = 1$. To fortify this claim, it would be interesting to explore the possibility of a positive energy theorem in the context of warped $AdS_3$ in TMG \cite{Witten:1981mf,Nester:1982tr,Sezgin:2009dj}. If stability indeed holds, one might suspect that an initial $AdS_3$ configuration in TMG, which is known to have perturbative instabilities, decays to stretched warped $AdS_3$.

We should point out that the boundary analysis in \cite{Compere:2008cv} only gave rise to a Virasoro extension of the right-moving isometry group, together with a $U(1)$ Ka\v{c}-Moody algebra. In particular, it did not give rise to the left moving, centrally extended, Virasoro proposed in \cite{Anninos:2008fx}. It would be very interesting if new consistent boundary conditions were found for warped $AdS_3$, which yield two copies of the Virasoro algebra as the ASG, with the expected central charges. Using the explicit negative-energy solutions found herein, it should be possible to immediately check the stability of warped $AdS_3$ with the new boundary conditions.

One of the puzzling features of global stretched $AdS_3$ is that it does not clearly have the same asymptotic structure as the stretched  $AdS_3$ black holes. It is unclear therefore what the meaning of a partition function for stretched $AdS_3$ is in this case, since one would like to sum over configurations with the same asymptotic behavior. It is possible that one would need to replace the global vacuum by the $r_+ = r_-=0$ black hole.

 We should also point out that stretched $AdS_3$ could still be unstable at a nonperturbative level.

Another open question is whether there exist boundary conditions under which the negative energy propagating solutions can be discarded in the squashed warped $AdS_3$ vacua. If we extrapolated the proposal in \eqref{compbc} we would conclude that the solutions are true instabilities. On the other hand, a different set of boundary conditions resembling those discussed in \cite{Guica:2008mu} may lead to an exclusion of all the propagating solutions for the squashed case as well.

We have also noted the propagating solutions exhibit some interesting properties in their own right. For instance there is a qualitative difference between the conformal weight of the stretched and squashed solutions. In particular, for the squashed solutions the conformal weight is real for all allowed values of $\nu$ and $k$, whereas in stretched $AdS_3$ it is only real for a small window of parameter space given by $k^2< \frac{(5-4a^2)a^2}{4(a^2-1)}$. When the weight becomes complex, the solutions propagate in the $r$-direction as well and thus there is a flux of energy escaping the boundary. This is highly reminiscent of the behavior of a scalar field in the near horizon geometry of the extremal Kerr black hole as was studied in \cite{Bardeen:1999px}. In fact, \cite{Bardeen:1999px} noted that this behavior was related to the superradiance of rotating black holes and we suspect a qualitatively similar phenomenon might be occurring in the stretched warped $AdS_3$ background, had we not discarded the solution.

It may also be worth noting that the mass squared of the massive gravitons \eqref{mass} becomes negative for the stretched case. Furthermore, when the warp factor satisfies $a^2 > 5/4$, the frequency of the highest weight solutions becomes complex for all values of $k$. It would thus be interesting to explore the behavior of the theory in these regimes.

Finally, we would like to point out that for the squashed case there is another candidate for a potentially stable vacuum given by timelike squashed $AdS_3$. This spacetime is related to spacelike squashed $AdS_3$ by an analytic continuation in the coordinates given by $\tau \to i x$ and $x \to - i \tau$. Thus our expressions for the highest weight solutions are closely related to the ones we have obtained. On the other hand, there are no black hole solutions or consistent boundary conditions known for this spacetime and it would be interesting to explore such questions. It would also be interesting to study the stability of null warped $AdS_3$ whose identifications also give rise to black holes.

\bigskip

\medskip

\noindent{\bf Acknowledgements}

\medskip
\noindent We would like to thank F. Denef, O. Dias, G. Giribet, T. Hartman, C. Keeler, H. Reall, W. Song and A. Strominger for illuminating discussions. We especially thank G. Comp\`ere and S. Detournay for useful discussions and checks of the boundary conditions. D.A. and M.G. would also like to thank the ESI for its kind hospitality while part of this work was completed. D.A. and M.E. have been partially funded by a DOE grant DE-FG02-91ER40654.

\appendix

\section{Properties of global warped $AdS_3$}

We review the geometry of global $AdS_3$ expressed as a Hopf fibration over $AdS_2$ and show geodesic completeness of spacelike warped $AdS_3$.

\subsection{Global vs. fibered coordinates in $AdS_3$}

The simplest way to picture $AdS_3$ is as a Lorentzian hyperboloid embedded in $\mathbb{R}^{2,2}$. If the coordinates on Minkowski space are $X^0, X^1, X^2, X^3$, then $AdS_3$ is the surface\footnote{We have set the $AdS_3$ radius $\ell=1$.}
\be
X_0^2 - X_1^2 - X_2^2 + X_3^2 =1
\ee
Different coordinate systems are obtained via different embeddings. To obtain $AdS_3$ in the usual global coordinates
\be
ds^2 = - \cosh^2 \rho\, dt^2 + d \rho^2 + \sinh^2 \rho\, d\phi^2 \label{ads3global}
\ee
with $\phi \sim \phi + 2 \pi$, we use the following parametrization
\bea
X^0 & = & \cosh \rho \cos t \;, \;\;\;\;\; X^1 = \sinh \rho \sin \phi \non \\
X^3 & = & \cosh \rho \sin t \;, \;\;\;\;\; X^2 = \sinh \rho \cos \phi
\eea
On the other hand, to obtain $AdS_3$ in the fibered coordinates
\be
ds^2 = \frac{1}{4} \left[ - \cosh^2 \sigma \, d\tau^2 + d\s^2 + (dx + \sinh \s\, d\tau )^2 \right] \label{ads3fabric}
\ee
we use the following embeddings
\bea
X^0 &=& \cos \frac{\tau}{2}\, \cosh \frac{x}{2}\, \cosh \frac{\sigma}{2} +
 \sin \frac{\tau}{2} \, \sinh \frac{x}{2} \, \sinh \frac{\sigma}{2} \non \\
 X^1 &=& \sin \frac{\tau}{2}\, \sinh \frac{x}{2}\, \cosh \frac{\sigma}{2} -
 \cos \frac{\tau}{2} \, \cosh \frac{x}{2} \, \sinh \frac{\sigma}{2} \non \\
 X^2 &=& - \cos \frac{\tau}{2}\, \sinh \frac{x}{2}\, \cosh \frac{\sigma}{2} -
 \sin \frac{\tau}{2} \, \cosh \frac{x}{2} \, \sinh \frac{\sigma}{2} \non \\
X^3 &=&- \sin \frac{\tau}{2}\, \cosh \frac{x}{2}\, \cosh \frac{\sigma}{2} +
 \cos \frac{\tau}{2} \, \sinh \frac{x}{2} \, \sinh \frac{\sigma}{2}
\eea
Note that in both cases we need to decompactify the time coordinate to avoid CTCs.

The coordinate transformation between the above coordinate systems is quite complicated, but with the help of the embeddings we immediately find
\be
-2(X^0 X^1 -X^2 X^3) =  \sinh \s =  \sin(t -\phi) \, \sinh 2 \rho \label{findsig}
\ee

\be
-2(X^0 X^2 + X^1 X^3) =  \cosh \s \, \sinh x = - \cos(t - \phi) \, \sinh 2 \rho \label{findx}
\ee
\bigskip
In the remainder of this section, we review the structure of the global $AdS_3$ boundary in the coordinates \eqref{sqads3} \cite{Bengtsson:2005zj,Balasubramanian:2003kq,Coussaert:1994tu}, in the  hope that they will help the reader improve his or her intuition about this kind of spaces.

In usual global coordinates \eqref{ads3global}, the boundary is at $\rho \r \infty$ and is famously a cylinder, parameterized by $t$ and $\phi$. We would like to find out where the boundary lies in terms of the fibered coordinates \eqref{ads3fabric} which are also global, but both $\tau$ and $x$ are noncompact.

\begin{figure}[h]
\centering
\includegraphics[width=4 cm]{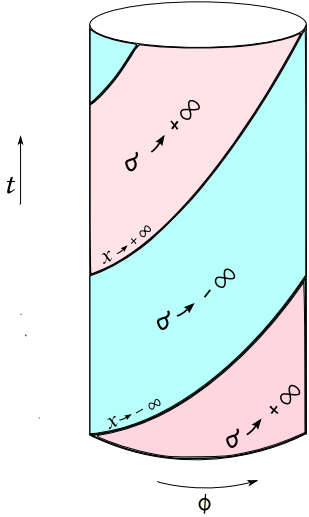}
\caption{A chromatic depiction of the way the fibered coordinates cover the $AdS_3$ boundary cylinder. Blue strips have $\s \r - \infty$ while pink strips have $\s \r + \infty$.} \label{adscyl}
\end{figure}

Using \eqref{findsig} we see that in order to reach the boundary at $\rho \r \infty$, we must take $\s \r \pm \infty$ whenever $t \neq \phi \; \mbox{mod} \; \pi$. Thus the boundary cylinder of $AdS_3$ is parsed by null strips of $\s \r \infty$ and $\s \r - \infty$, as shown in  figure \ref{adscyl}.

Next, we have to understand what happens on the null lines on the boundary $t = \phi  \; \mbox{mod} \; \pi$. Note first that all $\s =$ constant hypersurfaces end on these null lines on the boundary. From \eqref{findsig} and \eqref{findx} we conclude that
\be
\sinh x = - \frac{\cos(t - \phi) \, \sinh 2 \rho}{\sqrt{1+ \sin^2(t-\phi)\, \sinh^2 2 \rho}}
\ee
Thus, as we increase $t$ (with $\phi$ fixed), $x$ varies from $-\infty$ to $\infty$ on the pink ($\s \r \infty$) strips, while it varies from $\infty$ to $-\infty$ on the blue ($\s \r -\infty$) strips. Note that if we would like to go around the $\phi$ circle at fixed $t$ (say $0< t < \pi$), we first fix $\s \r \infty$ and take $x \r -\infty$, then fix $x$ at this value and take $\s \r -\infty$, then fix $\s$ and take $x$ from $-\infty$ to $\infty$, then cross back to the $\s \r \infty$ strip while keeping $x \r \infty$. Thus it is quite a bit more complicated to describe the compact direction on the $AdS_3$ boundary  in the fibered coordinates.

In conclusion, in terms of the fibered coordinates \eqref{ads3fabric}, the boundary of $AdS_3$ consists of the two {\it apparently} disconnected pieces at $\s \r \pm \infty$, but which are in fact connected at $x \r \pm \infty$ into the expected boundary circle.

\subsection{Geodesic completeness}

We review the geodesic completeness of warped $AdS_3$ \cite{Bardeen:1999px,Bengtsson:2005zj}\footnote{Geodesic completeness means that all geodesics are extendible to arbitrarily large positive and negative values of their affine parameter. In particular, boundary points are only reached at infinite affine parameter.}. Note that for $a^2 = 1$, our coordinates have been proven in \cite{Coussaert:1994tu} to be complete, i.e. they parameterize in a one-to-one fashion the full embedded hyperbola. It is also clear that the coordinates are global for $a = 0$.

The two conserved quantities associated to the Killing vectors $\xi_{(x)}^\mu  = \partial_x$ and $\xi_{(\tau)}^\mu = \partial_\tau$ are given by
\begin{eqnarray}
\frac{d x^\mu(\lambda)}{d\lambda}  \xi^{(x)}_\mu &=& g_{xx}\frac{ d x(\lambda) }{d\lambda} + g_{\tau x} \frac{ d\tau(\lambda) }{ d\lambda} = -p\\
\frac{d x^\mu(\lambda)}{d\lambda}  \xi^{(\tau)}_\mu &=& g_{\tau x}\frac{ d x(\lambda) }{d\lambda} + g_{\tau \tau} \frac{ d\tau(\lambda) }{ d\lambda} = -e
\end{eqnarray}
where $\lambda$ is the affine parameter and $p$ and $e$ are constants. The equation $ds^2 = 0$ determines the equation obeyed by $r(\lambda)$:
\be
p^2-e^2 a^2 +2 e p a^2  r(\lambda )-p^2 (-1+a^2 ) r(\lambda )^2 + L^4 a^2  r'(a )^2 = 0
\ee
where $L^2 \equiv \ell^2/(\nu^2+3)$. The null geodesic equation can be solved exactly for $r(\lambda)$. For $a^2 \neq 1$ and $p \neq 0$ we find in the limit of large $r$
\be
r(\lambda) \sim e^{\pm\frac{\sqrt{p^2(-1+a^2)}}{L^2 a} \lambda}.
\ee
Given that $r$ spans the whole real line, it is clear that the points at infinity are not reached for any finite value of the affine length. Also interesting is the fact that for $a^2 < 1$ the null geodesics carrying non-vanishing momentum along the $x$-direction are confined within $r < \infty$ \cite{Bengtsson:2005zj}.

We can also examine the equations obeyed by $\tau$ and $x$
\begin{eqnarray}
\tau(\lambda) &=& \frac{1}{B}\int_{\lambda_0}^\lambda d\eta \frac{\left(e - p r(\eta)\right)}{1+r(\eta)^2}\\
x(\lambda) &=& \frac{1}{L^2}\int_{\lambda_0}^\lambda d\eta \left(-\frac{p}{L^2 a^2} + \frac{r(\eta)(-e+p r(\eta))}{1+r(\eta)^2}\right)
\end{eqnarray}
Since there are no poles in the integrands, one sees that the infinities of $u$ and $\tau$ are not reached for any finite value of $\lambda$.

Timelike geodesics obey the equation $ds^2 = - d\lambda^2$, which leads to
\be
p^2-e^2 a^2 + L^2 a^2  +2 e p a^2  r(\lambda )+\left(-p^2 (-1+a^2 ) + L^2 a^2  \right) r(\lambda )^2 + L^4 a^2  r'(\lambda )^2 = 0
\ee
Once again, the solutions found asymptotically to be of the form
\be
r(\lambda) \sim e^{\pm\frac{\sqrt{p^2(-1+a^2)- L^2 a^2}}{L^2 a} \lambda}.
\ee
For $a^2 < 1$ the geodesics are confined. For $a^2 > 1$ we find that the geodesics can touch the boundary whenever $p^2 > L^2 a^2/(a^2-1)$, which is qualitatively different from the timelike geodesics in $AdS_3$. For the special value of $p^2 = L^2 a^2/(a^2-1)$ with $a^2 > 1$ our timelike geodesics soften to the form
\be
r(\lambda) \sim \lambda^2
\ee
In this case too, however, the boundary is reached only at infinite affine parameter. Thus, our spacetime is timelike and null geodesically complete.

\section{Gauge-fixing}

In this appendix we present a careful derivation of the gauge-fixing condition $h_{\mu x}=0$, which holds for all modes with $k \neq 0$.

Let us write the background 3d metric as
\be
ds^2 = g_{\mu\nu} d x^{\mu} d x^{\nu} + a^2 (d z + A_{\mu} dx^{\mu})
\ee
where for the rest of the section $\mu \in \{ 0,1\}$ and
\be
ds_2^2 = - (1+r^2)\, d \tau^2 + \frac{dr^2}{1+r^2} \;, \;\;\;\;\; A = r dt
\ee
The inverse metric reads
\be
g^{MN} = \left(\begin{array}{c|c} g^{\mu\nu} & - A^{\mu} \\ \hline - A^{\nu} & a^{-2} + A^2 \end{array} \right)
\ee
while the $3d$ Christoffel symbols are
\be
{}^{(3)}\Gamma^{\rho}_{\mu\nu} = {}^{(2)}\Gamma^{\rho}_{\mu\nu} + \frac{a^2}{2} g^{\rho\sigma} (A_{\nu} F_{\mu\sigma} + A_{\mu} F_{\nu\sigma}) \;,\;\;\;\;\; F_{\mu\nu} = \p_{\mu} A_{\nu} - \p_{\nu} A_{\mu} = L^2 \e_{\mu\nu}
\ee
\be
{}^{(3)}\Gamma^{z}_{\mu\nu} = - {}^{(2)}\Gamma^{\rho}_{\mu\nu} A_{\rho}- \frac{a^2}{2} A^{\sigma} (A_{\nu} F_{\mu\sigma} + A_{\mu} F_{\nu\sigma}) + \half (\p_{\mu} A_{\nu} + \p_{\nu} A_{\mu})
\ee
\be
\Gamma^{\lambda}_{z \mu} = \frac{a^2}{2} g^{\l \rho} F_{\mu \rho} \;, \;\;\;\;\;\Gamma^{z}_{z \mu} = \frac{a^2}{2} A^{\l} F_{\l \mu} \;, \;\;\;\;\;\Gamma^{\l}_{z z} = \Gamma^{z}_{z z} =0
\ee
Next, the strategy is as follows: we consider small perturbations of the background metric
\be
g_{MN} = \bar{g}_{MN} + h_{MN}
\ee
and expand them in Fourier modes in $z$
\be
h_{\mu\nu} (x,z) = \int d k \, h_{\mu\nu}^{(k)} (x) \, e^{i k z} \;, \;\;\;\;\; h_{\mu z} (x,z) = \int d k \, a_{\mu}^{(k)}(x) \, e^{ik z}\,  \non
\ee
\be
 h_{z z} (x,z) = \int d k \, \phi^{(k)}(x) \, e^{i k z}
\ee
We would now like to gauge-fix these perturbations. Under a general diffeomorphism, to leading order, the perturbation $h_{MN}$ transforms as
\be
\delta \, h_{MN} = \nabla_M \xi_{N} + \nabla_N \xi_{M}
\ee
The action of the above diffeomorphisms on the Fourier modes of the metric perturbation is
\be
\delta h_{\mu\nu}^{(k)} = \nabla_{\mu} \xi_{\nu}^{(k)} + \nabla_{\nu} \xi_{\mu}^{(k)} - a^2 (A_{\nu} F_{\mu\sigma} + A_{\mu} F_{\nu\sigma}) (\xi_{(k)}^{\sigma} - A^{\sigma} \xi_z^{(k)}) - (\nabla_{\mu} A_{\nu} + \nabla_{\nu} A_{\mu}) \xi_z^{(k)}
\ee
\be
\delta a_{\mu}^{(k)} = \p_{\mu} \xi_z^{(k)} + i k \,\xi_{\mu}^{(k)} - a^2 F_{\mu\l} (\xi^{\l}_{(k)} - A^{\l} \xi_z^{(k)}) \;, \;\;\;\;\; \delta \phi^{(k)} = 2 i k \,\xi_z^{(k)}
\ee
where $\xi_M^{(k)}$ are the Fourier modes of the diffeomorphisms. Now we turn to gauge fixing. It is quite easy to see that we can set
\be
\phi^{(k)} = a_{\mu}^{(k)} = 0 \;, \;\;\; k \neq 0
\ee
by fixing the corresponding modes of the diffeomorphisms. In this case, $h_{\mu\nu}^{(k)}$ is gauge-invariant.

We need to use a slightly different gauge-fixing for the zero-modes. $\phi^{(0)}$ is clearly gauge invariant, while $a_{\mu}^{(0)}$ can also be set to zero. The residual gauge transformations satisfy
\be
\xi^{\l}_{(0)} = A^{\l} \xi_z^{(0)} + \frac{1}{a^2 L^2} \,\e^{\l\mu}\, \p_{\mu}\xi_z^{(0)}
\ee
One can use this residual gauge freedom to set the trace of $h_{\mu\nu}^{(0)}$ to zero. Indeed
\be
\d h = 2 \nabla_{\mu} \xi^{\mu}_{(0)}- 2 a^2 A^{\mu} F_{\mu\nu} \xi^{\nu}_{(0)} - 2 \nabla_{\mu} A^{\mu} \xi_z^{(0)} = - 2 a^2 A^{\mu} F_{\mu\nu} \xi^{\nu}_{(0)}
\ee
which we can set to zero by choosing $\xi_{\mu}^{(0)} = \xi_z^{(0)} A_{\mu}$. All gauge freedom is fixed this way. We therefore conclude that a completely gauge-fixed form of the perturbation for $k \neq 0$ is
\be
h_{MN}(x^{\mu},k) = e^{i k z} \, \left( \begin{array}{c|c} h_{\mu\nu}^{(k)} & ~ 0 \\ \hline ~ 0 &~ 0 \end{array}\right)
\ee
while for $k=0$ we could have

\be
h_{MN}(x^{\mu},0) =  \left( \begin{array}{c|c} h_{\mu\nu}^{(0)} & ~ 0 \\ \hline ~ 0 &~ \phi^{(0)} \end{array}\right) \;, \;\;\;\;\; tr\, h^{(0)} =0
\ee
or alternatively use the gauge employed in \cite{Kim:2009xx}.

\section{Various expressions}

The expressions for the coefficients $C_{3,4}$ which enter the solution of the highest weight propagating modes read
\be
C_3 = \frac{(-1 + a^2) (-1 + \om) (-2 - a^2 + \om)C_5}{-2 +
  a^4 (-2 + \om) + \om - a^2 (-4 + \om + \om^2)}
\ee
\be
C_4 =
-\frac{a^2 \sqrt{-1 + a^2}\,
  (-2 + a^4 + \om (2 + (3 - 2 \om) \om) +
    a^2 (1 + \om (-5 + 2 \om)))C_5}{
 \sqrt{-a^2 (-1 + a^2 + (-1 + \om) \om)}\, (-2 +
    a^4 (-2 + \om) + \om -
    a^2 (-4 + \om + \om^2))}
 \ee
\bigskip
The expressions for the polynomials $P_{4,5,6}$ which enter into the linearized equations of motion for the propagating modes are
\bea
P_4(r)&= & a^4 \om^2 (a^6 - 4 a^2 k^2 + a^4 \om^2 +
    k^2 (-1 + \om^2)) + \non \\ &&
 2 a^4 k \om (a^6 - a^2 (1 + 4 k^2) + 2 a^4 (-1 + \om^2) +
    2 k^2 (1 + \om^2)) r + \non \\ && (a^2 (-1 + a^2) k^2 (-3 a^4 + a^6 +
       k^2 - a^2 (1 + 4 k^2)) +
    a^4 (a^6 - 4 a^2 k^2 + 6 k^4 + a^4 (1 + 6 k^2)) \om^2) r^2\non \\ && +
 2 a^4 k (a^2 (-1 - a^2 + a^4) + (3 - 4 a^2 + 2 a^4) k^2 +
    2 k^4) \om r^3\non \\ && +
 a^2 k^2 (-k^2 +
    a^2 (1 + 2 a^2 - 3 a^4 + a^6 + (6 - 4 a^2 + a^4) k^2 + k^4)) r^4
\eea
\bea
P_5(r) &=& -2 (a^4 k \om (a^6 - a^2 (1 + 4 k^2) + 2 a^4 (-1 + \om^2) +
       2 k^2 (1 + \om^2))) - \non \\ &&
 2 (a^2 ((-1 + a^2) k^2 (-3 a^4 + a^6 + k^2 - a^2 (1 + 4 k^2)) + \non \\ &&
      a^2 (a^4 - 2 a^6 + 3 k^2 + 8 a^2 k^2 + 6 a^4 k^2 +
         6 k^4) \om^2 - 3 a^2 (a^4 + k^2) \om^4)) r  \non \\ &&+
 2 a^4 k \om (2 a^6 - 2 a^2 (1 + 4 k^2) +
    a^4 (-7 - 6 k^2 + 10 \om^2) +
    k^2 (1 - 6 k^2 + 10 \om^2)) r^2 \non \\ &&+
 4 a^4 (-k^2 (1 + k^2) (a^4 + k^2) + (a^6 - 4 a^2 k^2 + 6 k^4 +
       a^4 (1 + 6 k^2)) \om^2) r^3\non \\ && +
 6 a^4 k (a^2 (-1 - a^2 + a^4) + (3 - 4 a^2 + 2 a^4) k^2 +
    2 k^4) \om r^4\non \\ && +
 2 a^2 k^2 (-k^2 +
    a^2 (1 + 2 a^2 - 3 a^4 + a^6 + (6 - 4 a^2 + a^4) k^2 + k^4)) r^5
\eea
\bea
P_6(r) &=&
a^2 \om^2 (a^{10} - k^4 (1 + \om^2) + a^8 (3 + 2 \om^2) -
     a^4 k^2 (3 + 4 \om]^2) + \non \\ &&
    a^2 k^2 (-1 + 2 k^2 + 2 \om^2 + \om^4) +
    a^6 (3 k^2 + 3 \om^2 + \om^4))\non \\ && +
 2 a^2 k \om ((-1 + a^2) (a^4 + k^2) (a^2 + a^4 + 2 k^2) +
    2 (-2 a^6 + 2 a^8 + 2 a^2 k^2 - k^4 - \non \\ &&
       a^4 (1 + 4 k^2)) \om^2 +
    3 a^2 (a^4 + k^2) \om^4) r \non \\ &&+ ((-1 + a^2) k^2 (a^4 + 4 a^6 -
       2 a^8 + a^{10} + a^4 (2 + 3 a^2) k^2 + (-1 + 2 a^2) k^4) + \non \\ &&
    a^2 (a^6 (-3 + 5 a^2 + 2 a^4) +
       2 a^4 (1 + 6 a^2 (-3 + a^2)) k^2 -
       2 (3 - 8 a^2 + 12 a^4) k^4 - 6 k^6) \om^2 + \non \\ &&
    a^2 (2 a^6 (4 + a^2) +
       a^2 (7 - 4 a^2 + 15 a^4) k^2 + (-1 +
          15 a^2) k^4) \om^4) r^2 \non \\ &&+
 2 a^2 k \om (2 a^{10} - 2 k^4 (4 + k^2 + \om^2) +
    a^8 (-1 + 4 k^2 + 4 \om^2) +
    a^4 (1 - 8 k^4 - 2 \om^2 + k^2 (10 - 8 \om^2))  \non \\ && +
    2 a^6 (3 (-1 + \om^2) + k^2 (-9 + 5 \om^2)) +
    2 a^2 k^2 (-3 + 7 \om^2 +
       k^2 (4 + 5 \om^2))) r^3 \non \\ &&+ (-k^2 (a^2 + k^2) (-2 k^2 +
       a^2 (2 + 4 a^2 - 7 a^4 + 7 a^6 -
          2 a^8 + (10 - 7 a^2 + 4 a^4) k^2 + k^4)) + \non \\ &&
    a^2 (a^6 (1 + a^2)^2 +
       a^2 (5 + 5 a^2 - 9 a^4 + 12 a^6) k^2 + \non \\ && (-5 + 44 a^2 - 24 a^4 +
          15 a^6) k^4 + 3 (-2 + 5 a^2) k^6) \om^2) r^4 +
 2 a^2 k (a^4 (-1 + a^2)^2 (1 + a^2) + \non \\ &&
    a^2 (-1 + 12 a^2 - 11 a^4 + 4 a^6) k^2 + (-6 + 16 a^2 - 8 a^4 +
       3 a^6) k^4 + (-2 + 3 a^2) k^6) \om r^5 \non \\ &&+ (-1 +
    a^2) k^2 (a^2 + k^2) (-k^2 +
    a^2 (1 + 2 a^2 - 3 a^4 + a^6 + (6 - 4 a^2 + a^4) k^2 + k^4)) r^6
\eea
The asymptotic form of the ratios of these polynomials that appear in the equations of motion is
\be
A(r) = \frac{2}{r} + \O(r^{-2}) \;, \;\;\;\;\; B(r)= \frac{(a^2-1)(a^2+k^2)}{a^2 r^2} + \O(r^{-3})
\ee

\section{Analysis of the linearized equation of motion}

In section 4.4 we have obtained the equation of motion for the propagating mode of TMG, which takes the following form
\be
\frac{d^2 w}{d z^2} + f(z) \frac{d w}{d z} + g(z)\, w =0 \label{geneqn}
\ee
The variable $z$ is real in our case, but it could in principle be complex. The above differential equation is said to have a \textit{regular singular point} at $z=z_0$ if $f(z), g(z)$ are not analytic at $z_0$, but $(z-z_0) f(z)$ and $(z-z_0)^2 g(z)$ are \cite{chow}.

If the differential equation only has regular singular points (or no singular points at all, which occurs when $f(z), g(z)$ are analytic in the whole domain of definition), then solutions to these equation can be constructed. If the singularities of $f$ and $g$ are worse than above then  the equation is said to have irregular singular points and oftentimes the solutions cannot be found.

In our case, as long as $P_{4}$ does not have roots of multiplicity more than one which do not occur concomitantly with roots of $P_5$, the differential equation has only regular singular points. The behavior of the solution in the neighborhood of such a point is given by (we set $z_0=0$ for simplicity)
\be
w(z) = z^{\a} \sum_{s=0}^{\infty} a_s z^s \label{solnwsing}
\ee
where $\a$ is called the exponent or index of the singularity and satisfies the equation
\be
\a (\a-1) + f_0 \a + g_0 =0 \label{indicialeq}
\ee
where $f_0, g_0$ are the constant terms in the Taylor expansion of $z f(z),z^2 g(z)$ around $z_0=0$
\be
z f(z) = \sum_{s=0}^{\infty} f_s z^s \;, \;\;\;\;\; z^2 g(z) = \sum_{s=0}^{\infty} g_s z^s \label{tyexpfg}
\ee
The expression \eqref{solnwsing} for the solution to the differential equation has radius of convergence equal to the radius of convergence of the Taylor expansions \eqref{tyexpfg}. There are two solutions, corresponding to the two roots of the indicial equation \eqref{indicialeq}. If the solutions coincide or differ by an integer, one of the solutions acquires a logarithmic term.

The regular singular point can also occur at infinity, in which case we change variables to $u = z^{-1}$. The equation \eqref{geneqn} can be rewritten as
\be
\frac{d^2 w}{d u^2} + p(u) \frac{d w}{d u} + q(u)\, w =0
\ee
with
\be
p(u) = 2 z - z^2 f(z) \;, \;\;\;\;\; q(u) = z^4 g(z)
\ee
The coefficients that now appear in the indicial equation are the constant terms in the expansion of
$2 - z f(z)$ and $z^2 g(z)$ as $z \r \infty$.  One can easily show that if $\a$ is a solution of the following indicial equation
\be
\a (\a +1) - p_0 \a + q_0 =0
\ee
where
\be
z f(z) = \sum_{s=0}^{\infty} \frac{p_s}{z^s} \;, \;\;\;\;\; z^2 g(z) = \sum_{0}^{\infty}\frac{q_s}{z^s}
\ee
then the solution asymptotically takes the form
\be
w(z) = \frac{1}{z^{\a}} \sum_{s=0}^{\infty} \frac{a_s}{z^s}
\ee
In conclusion, what must be done is to find the zeroes of $P_4(r)$, make sure they do not lead to poles of multiplicity higher than one in $A(r)$, and match the solutions on the different patches.

\section{Consistency checks of the boundary conditions}

As reviewed in section 3.3, the conserved charges in the Barnich-Brandt formalism \cite{Barnich:2001jy} are constructed as boundary integrals of a given one-form, $K_{\mu}$ which is determined by the equations of motion of the theory. For TMG, the appropriate form of $K_{\mu} = \e_{\mu\nu\rho} F^{\nu \rho}$ was given in \cite{Compere:2008cv}. In this section, we consider the charges 
\be
Q_{\xi} = \int_{\p \Sigma} K_{\mu} dx^{\mu},
\ee
where $\p \Sigma$ is the boundary of global warped $AdS_3$. We reduce the questions of their integrability, finiteness and conservation to the question of consistency of the boundary conditions for the black hole spacetimes \eqref{ng}. The latter is answered in the affirmative in \cite{cdwip}. Since we do not know what precise contour the boundary circle of global warped $AdS_3$ follows, our strategy is to analyze each component of the integrand $K_{\mu}$. While we do restrict ourselves to linear perturbations around global warped $AdS_3$, we believe that the simple structure we find also holds for perturbations around any background allowed by the boundary conditions \eqref{compbc}.

The asymptotic symmetry group is generated by diffeomorphisms of the form
\be
\xi^{\tau} = f(\tau) \;, \;\;\;\; \xi^{r} = - r f'(\tau) \;, \;\;\;\; \xi^{x} = F(\tau)
\ee
Let $E_{\mu\nu}(\tau,x)$ be the linearized equations of motion for the asymptotic perturbation $h_{\mu\nu}$, and let $E^{(x)}_{\mu\nu} (\tau,x)$ be its indefinite integral with respect to $x$: $  \p_x E^{(x)}_{\mu\nu}(\tau,x)=E_{\mu\nu}(\tau,x)$. We find the following very simple structure
\bea
K_{\tau}& = & r \cdot f(\tau) E^{(x)}_{\tau r}  + \mathcal{O}(r^0)\non \\
 K_{x} & = & f(\tau) E_{rr} + \p_x (\mbox{something}) +  \mathcal{O}(r^{-1})\non \\
 K_{r} & = & \mathcal{O}(r^{-1})
\eea
The equations of motion imply that the divergent part of $K_{\tau}$ is simply an integration constant which is in fact zero \cite{cdwip}.

If we impose restrictive enough boundary conditions at large $x$, it is clear that $K_x$ and $K_r$ will in fact vanish since they only get contributions from $|x| \r \infty$, which are eliminated by our unspecified boundary conditions. The expression for the charges is then identical to the warped black hole case, except for the range of the $\tau$ integration. Consistency of our boundary conditions immediately follows, since the charges have been proven to be finite, integrable and conserved at the full nonlinear level for the warped black hole case.

This analysis strongly points towards the consistency of the boundary conditions \eqref{compbc} for an appropriate falloff at large $|x|$. However, one must still prove the integrability of the charges and show that similar simplifications occur for the full asymptotic equations of motion. Also, it is not clear how restrictive the boundary conditions in $|x|$ can be. In particular, it is not a priori clear whether wave packets are allowed by our boundary conditions for all time: if we start with a wave packet localized in the $|x|$ direction, it may spread to large $x$ in time\footnote{We are grateful to G. Comp\`ere for pointing this out.}. While wave packets in stretched $AdS_3$ are already excluded by their $r$-falloff, it would be nice to better understand the boundary conditions at large $|x|$.

\end{document}